# Vectorial Symmetry Decoding with Single-Particle Precision via Room-Temperature Lanthanide Luminescence Polarimetry

Peng Li[1,2]†*, Yaxin Guo[2]†, Yaoxu Yan[3], Bingzhu Zheng[4], Wenchao Zhang[5], Jingai Mu[2], Fu Liu[2]*, Yanpeng Zhang[2], Feng Yun[2,6], Rongqian Wu[1]*, Yi Lyu[1], Renren Deng[5]*, Feng Li[2,6]*

[1]Institute of Regenerative and Reconstructive Medicine, Med-X Institute, The First Affiliated Hospital of Xi'an Jiaotong University, Xi'an 710061, China

[2]Key Laboratory for Physical Electronics and Devices of the Ministry of Education and Shaanxi Key Lab of Information Photonic Technique, School of Electronic Science and Engineering, Faculty of Electronic and Information Engineering, Xi'an Jiaotong University, Xi'an 710049, China

[3]State Key Laboratory of Modern Optical Instrumentation, CNERC for Optical Instruments, Zhejiang University, 38 Zheda Road, Xihu District, Hangzhou 310027, China

[4]Department of Medical Oncology, The First Affiliated Hospital, School of Medicine, Zhejiang University, Hangzhou 310003, China

[5]State Key Laboratory of Silicon and Advanced Semiconductor Materials, Institute for Composites Science Innovation, School of Materials Science and Engineering, Zhejiang University, Hangzhou 310058, China

[6]Solid-State Lighting Engineering Research Center, Xi'an Jiaotong University, Xi'an 710049, China

†These authors contributed equally: Peng Li, Yaxin Guo

*Correspondence to: ponylee@xjtu.edu.cn, fu.liu@xjtu.edu.cn, rwu001@mail.xjtu.edu.cn, rdeng@zju.edu.cn, felix831204@xjtu.edu.cn

**Abstract:**

Determining the local symmetry of luminescent centers in crystals is critical for understanding and controlling their optical transitions, yet current methods are limited by stringent experimental requirements and ambiguous symmetry assignments. Here, we develop a robust computational electromagnetics framework that directly connect the local symmetry and chirality of rare-earth-doped single crystals to the polarization states of their emitted light. This framework is experimentally validated through the precise determination of point and space group symmetries using high-resolution, polarization-resolved micro-photoluminescence (μ-PL) spectra. Unlike conventional approaches that usually rely on analyzing multiple transitions at cryogenic temperatures, our technique operates at room temperature, requires only a single optical transition, and enables accurate orientation of symmetry axes. This enables deterministic polarization control of nano-emitters by tailoring symmetry groups and selecting appropriate transition dipoles, eliminating the need for bulky or complex photonic structures. Additionally, we demonstrate the function of bio-sensing, via determining single particle orientations in complex cellular environments using minimal polarization measurements. These results pave the way for advances in energy transfer systems, ultra-bright rare-earth nanocrystals, nanophotonic materials, and real-time single-particle tracking in biological contexts.

**Introduction**

Symmetry principles are fundamental to human civilization, underpinning key advances in physics and crystallography[1-4]. In particular, the local point group symmetry exerts a profound influence on the optoelectronic properties of luminescent centers in solids[5-10], playing a critical role across a wide range of applications including displays[11-14], lasers[15-17], super-resolution imaging[18-25], ultrasensitive measurements[26-32], and quantum photonics[33-39]. While the physical properties of crystals are intrinsically linked to the symmetry groups they possess, this relationship becomes significantly more complex when the crystals are doped with luminescent centers. The introduction of doped ions often leads to atomic-scale symmetry breaking[40], which is difficult to predict theoretically and cannot be easily detected by traditional characterization techniques such as X-ray or neutron diffraction[41]. Specifically, the point group symmetry of doped ions often deviates markedly from the undoped crystals, resulting in substantial changes to their optical properties. This symmetry breaking has a pronounced effect on the optical transitions of rare-earth ions, as evidenced by spectral splitting observed in cryogenic high-resolution spectroscopy[42]. While the type of point group can be inferred from the measured spectral splitting, determining the vectorial symmetry—specifically the orientation of the symmetry axis—remains a considerable challenge. Meanwhile, the recently developed computational approach for indexing symmetry axes requires atomic-scale information of coordinates which is hard to obtain in practice[43]. Moreover, an efficient method for precisely determining the space group is still lacking. To gain a comprehensive understanding of light-emitting mechanisms in doped solids, a practical and reliable method is needed to determine both the point group and the orientation of the symmetry axis, as well as the space groups in luminescent centers. This is crucial for the design of advanced optoelectronic devices through purposeful tailoring of light emission.

In recent years, it has been demonstrated that the polarization properties of emitted light from rare-earth-doped materials are strongly linked to the orientation of transition dipoles[31,44], and thus to the symmetry of the doped crystal[45]. Although polarized emission is typically averaged out in ensemble micro- or nanocrystals due to the random orientation of the crystals, recent advances in single-particle spectroscopy have enabled direct measurement of polarized emission in micro- and nanocrystals[31,44-54], opening up new applications in microfluidics and biological sensing[30-32]. It is highly anticipated that full information regarding the vectorial symmetry can be obtained through polarization-resolved micro-spectroscopy. However, a rigorous and quantitative framework that links light polarization, transition dipole orientation, and point group symmetry—incorporating insights from solid-state physics, advanced electromagnetics, and group theory—has yet to be fully developed.

In this study, we introduce a universal model system based on a rigorous computational electromagnetic framework that quantitatively links far-field polarization, transition dipole orientation, and point group symmetry. Through the polarization mapping of a single optical transition at room temperature, we reliably determine the point group of the doped emitter with full information about the symmetry axes, representing a significant advantage over traditional methods which require low-temperature measurements of multiple transitions. The validity of the method is experimentally confirmed through precise polarization-resolved micro-spectroscopy of the $^5D_0 \rightarrow {}^7F_1$ optical transition in $Eu^{3+}$-doped hexagonal yttrium phosphate ($YPO_4$) single microcrystals. We also observe opposite optical chirality in the single particles, linked to spontaneous symmetry breaking of chiral space group which cannot be determined by conventional approaches. Our findings demonstrate that optical polarization can be tuned by selectively choosing specific dipole transitions associated with a designed symmetry groups, offering a distinctive method for polarization control by leveraging the intrinsic symmetry of emitters, rather than relying on complex optical structures.

**Results and Discussion**

The basic logic of the model system, as illustrated in **Fig. 1**, consists of four successive steps. (1) A single particle of ion-doped microcrystal is experimentally measured through polarization-resolved micro-photoluminescence (µ-PL), from which the full polarization information of the desired optical transition is extracted. (2) The core model based on computational electromagnetics is then applied to link the measured linear polarization degree to the orientation of the transition dipoles, determining the orientations of the electric or magnetic dipoles and rotors within an optical transition. (3) The point group, which directly characterizes local symmetry breaking, is robustly determined from the orientations of the transition dipoles and rotors, based on the distinct causal relationship between them. Additionally, the circular polarization degree of light reveals the optical chirality associated with the space group of the crystal. (4) The obtained symmetry properties are then used to predict and design new structures for radiation engineering and polarization control. While the methodology is universal, we begin with the specific example of hexagonal $YPO_4$:$Eu^{3+}$ microcrystals to provide a clearer description of the underlying mechanism.

The investigated $YPO_4$:$Eu^{3+}$ microcrystals are approximately 2.5 µm in length and 1.5 µm in diameter (**Supplementary Fig. 1**), while the approach is generally applicable to single crystals of any size. For these microcrystals, the unit cell (**Fig. 2a**) consists of three $Y^{3+}$ layers arranged in an ABCA stacking sequence along the c-axis, which defines the long axis of the microcrystals. Projections of the $Y^{3+}$ sites form a Kagome lattice with sixfold rotational symmetry in the x-y plane (**Fig. 2b**), and partial or complete substitution of $Y^{3+}$ by doped $Eu^{3+}$ ions preserves this symmetry. Now we focus on the magnetic dipole transitions of the $Eu^{3+}$ ions under non-resonant excitation, where the orientations of the dipoles defined by their polar angles $\alpha$ relative to the c-axis. The uniformity and symmetry of the crystal ensure that all magnetic dipoles share the same polar angle $\alpha$ but have different azimuthal orientations, as illustrated in **Fig. 2c**. Under far-field conditions, the collective behavior of all the dipoles is equivalent to a group of six large dipoles arranged in six-fold rotational symmetry, as determined by the hexagonal structure of the crystal. The total light emission is the integration of the far-field electromagnetic fields of these six large dipoles. It is important to note that this integration is an incoherent superposition of light intensities rather than electric field amplitudes, as spontaneous emission from the excited ions does not induce phase coherence. Therefore, while each dipole behaves as a source of purely linear polarization, the incoherent integration of dipole emissions with different azimuthal orientations leads to partial polarization, characterized by a polarization degree less than one. By adopting a Cartesian coordinate system and decomposing the electric field of light into three polarization components ($E_x$, $E_y$ and $E_z$), we derive analytical expressions for the total polarized intensity from six incoherent dipoles (see **Supplementary II-A** and **II-B** for mathematical details):

$$I_x(\alpha, \theta, \varphi) = |E_x|^2 = \frac{3}{r^2}(2\cos^2\alpha \sin^2\theta \sin^2\varphi + \sin^2\alpha \cos^2\theta) \tag{1}$$

$$I_y(\alpha, \theta, \varphi) = |E_y|^2 = \frac{3}{r^2}(2\cos^2\alpha \sin^2\theta \cos^2\varphi + \sin^2\alpha \cos^2\theta) \tag{2}$$

$$I_z(\alpha, \theta, \varphi) = |E_z|^2 = \frac{3}{r^2}\sin^2\alpha \sin^2\theta \tag{3}$$

in which $\vec{r}$ ($\theta$, $\varphi$) represents the polar coordinates of the same system depicted in the inset of **Fig. 2c**. It is important to emphasize that, since the individual dipoles are incoherent, there is no phase correlation between $E_x$, $E_y$ and $E_z$. The equations yield $I_x(\theta = 0) = I_y(\theta = 0) = I_z(\theta = \pi/2)$, meaning that the intensities of the linearly polarized components emitted along the axial direction are equal to those along the radial direction, confirming the nature of magnetic dipole transitions in hexagonal crystals[30,45]. It is noteworthy from **Eqs. 1-3** that, generally, the polarization of light varies continuously with the detection direction ($\theta$, $\varphi$), meaning that polarization properties cannot be discussed without specifying the detection direction. Specifically, when detecting along the z-axis (axial detection), $\theta = 0$ yields $I_z = 0$ and $I_x = I_y$, indicating the emission of completely unpolarized light, as reported in numerous experimental observations[45,48,49,54]. On the other hand, for radial detection (i.e., detection in the x-y plane, as in the experimental setup used in this work and many others), the condition $\theta = \pi/2$ results in $I_z/I_T = (\tan^2\alpha)/2$, where $I_T = I_x + I_y$ is the intensity of the linearly polarized field component perpendicular to the z-axis. This quantitative relation indicates partial polarization that is distinctly linked to the dipole orientation $\alpha$. In the extreme case of $\alpha = 0°$ (resp. $\alpha = 90°$), pure linear polarization perpendicular (resp. parallel) to the z-axis is derived with $I_z = 0$ (resp. $I_T = 0$). However, in reality, $\alpha$ typically falls between 0° and 90°, and different magnetic dipoles—originating from fine spectral splitting of the same optical transition due to local symmetry breaking—can exhibit different values of $\alpha$, leading to dramatically different polarization properties quantified by $I_z/I_T$. In particular, when $\alpha = 54.7°$, the system reaches completely unpolarized emission, characterized by $I_T = I_z$.

To further demonstrate the advantage of the universal model of computational electromagnetics, we present the visualization of the calculated field distribution of the six-fold symmetric dipoles, along with the integrated emission, in **Fig. 2d** for the case where $\alpha = 54.7°$. From the integrated emission (upper right inset), it is observed that $I_T = I_z$ holds true only within the x-y plane ($\theta = \pi/2$), while $I_z/I_T$ is, in fact, completely anisotropic with respect to $\theta$. Such accurate angular-dependent information is essential for a comprehensive understanding of the mechanisms of light-matter interaction. Although a similar expression for $I_z/I_T$ was derived in our previous study by coarsely approximating dipole emission as planar waves[45], only the rigorous approach of computational electromagnetics in this work quantitatively predicts the polarization anisotropy, which precisely reflects the nature of dipoles. Similar visualizations for other values of $\alpha$ are provided in **Supplementary Fig. 5**. For clarity, the variation in polarization with $\alpha$, specifically for radial detection (calculated herein along the x-axis), is plotted graphically in **Fig. 2e** and statistically in **Fig. 2f**. The plot predicts the following scenario: the degree of linear polarization decreases with $\alpha$ until reaching zero at $\alpha = 54.7°$, after which it increases with $\alpha$, accompanied by a 90° rotation of the polarization angle. Indeed, deriving a mathematical expression for the relationship between the degree of linear polarization (LDOP) and $\alpha$ is straightforward, as LDOP is closely related to $I_z/I_T$. As analytically deduced, polarization angle must be either parallel or perpendicular to the z-axis (see **Supplementary II-C, Eq. S22**). Thus, we have $LDOP = |I_z - I_T|/(I_z + I_T)$, and therefore

$$(1 \pm LDOP)/(1 \mp LDOP) = (tan^2\alpha)/2 \qquad (4)$$

by considering $I_z/I_T = (tan^2\alpha)/2$. Herein + (resp. −) in the numerator corresponds to polarization angle parallel (resp. perpendicular) to the z-axis. With this quantitative relation, we successfully link the linear polarization angle and degree to the orientation of the transition dipole, thus have completed the theoretical model required in step (2) depicted in **Fig. 1**.

We now describe the experimental methodology for determining the LDOPs of the emission peaks, linking step (1) to step (2). Experimentally, we selected the $^5D_0 \rightarrow {^7F_1}$ optical transition for investigation, as it is split into three resolvable emission peaks (denoted as peaks I, II, and III) at room temperature due to point group symmetry breaking (**Supplementary III-B, Fig. 9**). Spectral resolution of individual emission peaks is essential for this analysis, as the corresponding magnetic dipole transitions (MD I, MD II, and MD III) can be fitted with our theoretical model. The μ-PL measurements were performed at room temperature; technical details are provided in the **Methods** section. A microscopic image of the measured YPO$_4$:Eu$^{3+}$ microcrystal is shown in **the inset of Fig. 3a**. The c-axis (z-axis in **Fig. 2**) lies parallel to the substrate, forming an angle $\Omega_c \approx 156°$ relative to the lab coordinate frame. Herein we link the experimental coordinate with the theoretical model by denoting $I_\parallel \equiv I_z$ and $I_\perp \equiv I_T$, in which ∥ (resp. ⊥) mean parallel (resp. perpendicular) to the c-axis of the microcrystal which exhibits an angle $\Omega_c$ with respect to the defined 0° of the lab coordinate. Polarization-resolved measurements were acquired in the lab coordinate system for linear polarizations (0°, 90°, ±45°) and circular polarizations (right-hand, σ$^+$; left-hand, σ$^-$); the data are presented in **Fig. 3a**. From these measurements, we computed the Stokes parameters $S_1$, $S_2$ and $S_3$ (See **Supplementary III-A, Fig. 7** for detail) as functions of wavelength (**Fig. 3b**). The emission spectrum comprises three Lorentzian-fitted peaks, as shown in **Fig. 3c**. The Stokes parameters were then calculated from the integrated intensity of each peak, yielding ($S_1$, $S_2$, $S_3$) for peaks I–III as (−0.38, 0.41, 0.02), (−0.06, 0.05, −0.01), and (0.49, −0.63, 0.08), respectively. Since the circular polarization degree ($S_3$) is near zero—consistent with dipole emission—all peaks lie near the equatorial plane of the Poincaré sphere (**Fig. 3d,** see also **Supplementary III-A, Fig. 8** for an introduction of the Poincaré sphere). The linear polarization angles and degrees can thus be derived by $\Omega = (arctan(S_2/S_1))/2$ and LDOP = $\sqrt{S_1^2 + S_2^2}$. This yields $\Omega$: 66.3°, 70.2°, 154.0° and LDOPs: 0.55, 0.07, 0.80 for peak I to III.

In addition to Poincaré sphere (PS) measurements, we performed polarization fitting (PF) to validate the accuracy of our derivation (see **Supplementary II-C, Eq. S23** and **Supplementary III-A, Fig. 7** for details). For PF analysis, we measured the emission spectrum at various detection angles $\omega$ of the polarization analyzer, as shown in **Fig. 3e**. By fitting the measured intensity of each peak using: $I_\omega = (A - B)cos^2(\omega - \Omega) + B$ (or equivalently $I_\omega = Acos^2(\omega - \Omega) + Bsin^2(\omega - \Omega)$), we derived the linear polarization angle $\Omega$ and the LDOP = $(A - B)/(A + B)$. The PF results are visualized in **Fig. 3f**, yielding $\Omega$: 66.4°, 66.4°, 156.4° and LDOPs: 0.54, 0.08, 0.77 for peaks I–III, respectively. These results are nearly identical to those obtained from PS analysis. The derived data reveal that the three magnetic dipoles, despite originating from the same $^5D_0 \rightarrow ^7F_1$ optical transition, exhibit dramatically different emission polarizations. First, MD I and MD III emit orthogonally polarized photons: peak I is polarized perpendicular to the c-axis ($\Omega= 66.4°$ versus $\Omega_c \approx 156°$), while peak III is polarized parallel to the c-axis ($\Omega= 156.4°$). Second, substantial differences in LDOP are observed among the three dipoles. Notably, MD II emits nearly unpolarized photons with an LDOP near zero (0.07–0.08), resembling a dipole in vacuum unaffected by the crystal field. We verified the robustness of our polarization-resolved measurements by analyzing nine additional YPO$_4$:Eu$^{3+}$ microcrystals with different in-plane orientations (**Supplementary Figs. 11,12**). All microcrystals exhibited highly consistent partial linear polarization with a standard deviation < 0.05 in LDOP (**Supplementary Tables 2,3**). Using the mathematical relation in **Eq. 4**, we obtained magnetic dipole orientations of $\alpha_I= 37.6°$ ($-37.6°$), $\alpha_{II}= 52.8°$ ($-52.8°$), and $\alpha_{III}= 75.3°$ ($-75.3°$) for MD I–III. Both the positive and negative angles, being mirror images of each other, result in the same situation. The spatial relationship of these dipoles is illustrated in **Fig. 3g**, and statistical results from ten measured microcrystal samples are shown in the **left panel of Fig. 3h**. Having completed steps (1) and (2) depicted in **Fig. 1**, we now proceed to step (3).

Step (3) determines the point group of the doped ions based on the obtained dipole orientations. We begin by examining all 32 crystallographic point groups (**Supplementary III-B, Fig. 10, and Table 1**). The hexagonal YPO$_4$ microcrystal intrinsically exhibits D$_6$ point group symmetry for Y$^{3+}$ ions[9,55-58]. However, local symmetry breaking induced by Eu$^{3+}$ doping could theoretically lead to decomposition toward D$_2$, C$_6$, and D$_3$, or further toward C$_2$, C$_3$, and C$_1$ point groups (**Supplementary Fig. 10**). We then evaluated the selection rules for the $^5D_0 \rightarrow {}^7F_1$ optical transition in all six point groups (**Supplementary Table 1**) to determine which agrees with our Step (2) results. The D$_2$ symmetry produces three emission peaks corresponding to mutually orthogonal magnetic dipoles R$_X$, R$_Y$, R$_Z$, satisfying:

$$\cos\alpha_I + \cos\alpha_{II} + \cos\alpha_{III} = 1 \qquad (5)$$

This relation can be tested against our experimental data. Substituting the measured values for MD I ($\alpha_I= 37.6°$) and MD III ($\alpha_{III}= 75.3°$) into **Eq. 5**, we calculated the orientation of MD II as $\alpha'_{II}= 56.3°$, which agrees well with the measured value $\alpha_{II}= 52.8°$. This verification was performed on ten different microcrystals (**Supplementary Table 4**), with statistical results shown in the **right panel of Fig. 3h**. The small discrepancy of 3.5° between calculated ($\alpha'_{II}$) and measured ($\alpha_{II}$) values, within experimental error, arises from measurement inaccuracies associated with light collection at non-zero angles to the x-y plane and non-zero circular polarization generated during light propagation through the crystal (discussed later). Other point groups can be excluded as they exhibit either fewer than three transition peaks ($C_6$, $D_3$, $C_3$) or non-orthogonally oriented dipoles or rotors ($C_2$, $C_1$). Therefore, we robustly determine the point group of rare-earth dopants (here, $Eu^{3+}$) as $D_2$, consistent with crystal field theory calculations[57]. Importantly, the vectorial properties of the $D_2$ point group—the three orthogonal $C_2$ axes defining the symmetry invariance under 180° rotation operations—are simultaneously determined. These $C_2$ axes correspond to the orientations of $R_X$, $R_Y$, $R_Z$ illustrated in **Fig. 3g**, which occur to be the orientations of the magnetic dipoles for $D_2$ point group (**Supplementary Table 1**). It is noteworthy that the experimentally determining the exact orientations of symmetry axes for a given point group has been exceptionally challenging. Indeed, the $C_2$ axes revealed here do not align with the c-axis of the microcrystal, making them inaccessible by conventional methods such as crystallographic diffraction or polarization-unresolved spectral measurements. Another advantage of our method is that it requires measurement of only a single optical transition at room temperature. In contrast, conventional low-temperature spectroscopic methods require numerous transitions at cryogenic temperatures[40,42], yet fail to provide vectorial information.

It should be noted that the computational electromagnetic method is universal. While it is applicable to dipoles, it also extends to rotors (combinations of two orthogonal dipoles). In addition to magnetic transitions, the method is also applicable to electric optical transitions, where the mathematical expressions for electric dipoles and rotors vary from **Eq. 1** to **Eq. 3**. Particularly in cases where two or more point groups are difficult to distinguish based on magnetic transitions (e.g., $C_6$ vs. $D_3$), an additional measurement of an appropriate electric transition can resolve the issue. In principle, the concept of constructing the model system applies to all crystal systems, not only for the hexagonal one. Although computational electromagnetics may not yield analytical expressions for low-symmetry crystals (e.g., monoclinic), problems can still be solved through numerical calculations, by plotting the calculated field distributions as in **Fig. 2d**.

It is noteworthy that a non-negligible degree of circular polarization ($S_3$) for peak III is observed for individual microcrystals, as statistically shown in **Fig. 4a** for the ten samples measured in **Fig. 3a** and **Supplementary Fig. 11**. The nonzero $S_3$ is not an intrinsic property of the local point group symmetry breaking associated with dipole transitions; rather, it emerges during light propagation within the microcrystal, reflecting the crystal's chirality. Examination of $S_3$ reveals six right-handed ($S_3 > 0$) and four left-handed ($S_3 < 0$) microcrystals among the ten samples, with an average $S_3 = 0$, indicating a racemic ensemble. This is further confirmed by the ensemble measurements presented in **Figs. 4b,c**. The coexistence of left and right circular polarization indicates two space groups of opposite chiralities. While hexagonal YPO$_4$ is known to form in the chiral P6$_2$22 space group[55-58], its enantiomorph P6$_4$22 exhibits the opposite chirality. Although conventional X-ray diffraction[59] cannot tell the exact space group of the doped crystal as P6$_2$22 and P6$_4$22 share identical diffraction pattern (**Supplementary Fig. 1**), our polarization-resolved μ-PL confirms their coexistence with spontaneous stochastic distribution, revealing spontaneous chiral symmetry breaking during crystallization. By identifying both the point group and space group of the system, we complete step (3) outlined in **Fig. 1.**

The intrinsic relationship between the crystal symmetries and the light polarization enables efficient polarization control through the design of crystal structures, in stark contrast to the conventional approach that relies on external optical structures, such as optical cavities or metamaterials, which require complex fabrication and large space for samples. Indeed, Peak II of the $^5D_0 \rightarrow ^7F_1$ transition under the D$_2$ point group (**Fig. 3f**) is a clear example. When detected along radial directions, it exhibits unpolarized emission, as if the crystal field is completely screened, creating a vacuum-like emitting environment. The emergence of such performance should not be viewed as a coincidence, as there is a vast array of point and space groups associated with each crystal system, an extensive range of electric and magnetic optical transitions corresponding to each point group, and a broad spectrum of rare-earth ion types covering the entire emission range. Given this abundance of options, it is reasonable to believe that, except in extreme cases (e.g., LDOP = 1), most polarization states can be achieved by designing an appropriate ion-doped crystal and selecting the correct optical transition. This completes step (4) in **Fig. 1.**

Another important application of polarization-resolved μ-PL is the direct determination of the 3D orientation of single particles at room temperature, owing to the anisotropy of the emitted polarization. This orientation determination is particularly useful for bio-sensing and has been demonstrated by analyzing several transition peaks[45]. Herein, we develop an improved method that allows for the determination of the 3D orientation by analyzing the polarization of transition bands, rather than relying on well-resolved transition peaks associated with individual dipole transitions. This band-based analysis eliminates the need to resolve single emission peaks, making it especially suitable for sensing applications under ambient conditions. To experimentally demonstrate the concept, we still choose the $^5D_0 \rightarrow {}^7F_1$ transition for analysis. Instead of fitting the individual peaks, we divide the transition into two bands (denoted as band I and II) and perform the polarization-resolved µ-PL as presented in **Fig. 5a-d**, with the statistical results of linear polarization degrees (LDOPs) for ten samples in **Fig. 5e**, displaying a standard deviation less than 0.02 (**Supplementary Table 5**). The measurements show that band I (resp. band II) shows partial linear polarizations perpendicular (resp. parallel) to the crystal c-axis. The principle of 3D orientation detection is illustrated in **Fig. 5f**. In this approach, we employ an orthogonal linear polarization detection scheme within a well-defined Cartesian coordinate system, capturing the relative intensities of the two transition bands under orthogonal polarizations. By combining these intensity ratios with the characteristic LDOPs of each band, we can unambiguously determine the 3D orientation of individual particles. All mathematical relationships and details of the extraction procedure are provided in **Supplementary IV**.

We further validate the proposed method by performing bio-sensing with single particles in complex cellular environments. Single microcrystal particles are incorporated into a cell, as microscopically shown in **Fig. 5g**, resulting in randomized distributions of 3D orientations. Through analysis based on orthogonally-polarized spectroscopy along a single direction of view (or detection), we derive the in-plane orientation angles of two typical particles as 17.3° and 172.4° (**Supplementary Fig. 13 and Supplementary Table 6**). Direct measurements from the microscopic image (**Fig. 5h,i**) identify the in-plane angles as 16° and 170°, in very good agreement with the polarization analysis. Interestingly, the polarization analysis also reveals tilt angles of 14.2° and 19.2° for the two particles with respect to the substrate, corresponding to the gravitational alignment of the surrounding environment, a feature not directly observed in the microscopic images. These results highlight the potential of single-particle polarimetry to resolve local structural environments within complex biological systems, such as cellular architectures.

**Conclusion**

We developed a computational electromagnetics framework that enables the determination of point group symmetry with full vectorial properties using polarization-resolved spectroscopy at room temperature, surpassing traditional low-temperature and crystallographic methods. This approach facilitates analysis of system chirality, symmetry-based polarization control, and bio-sensing of particle orientations in complex cellular environments, potentially enabling real-time tracking of nanoscale dynamics such as protein rotation, organelle organization, DNA conformational changes, and cytoskeletal reorganization. It can also be used to predict extreme polarization phenomena—including superfluorescence[60], superradiance, and stimulated emission amplification—arising from coherent dipole alignment in single rare-earth emitters. Additionally, the framework allows quantitative evaluation of the Förster dipole orientation factor ($\kappa^2$)[61,62], aiding the optimization of donor–acceptor arrangements and the design of ultra-bright rare-earth nanocrystals. Our findings open new avenues for optical tools to probe structure and dynamics in single nanoscale systems, advancing discoveries in materials and life sciences.

**Methods**

**Single-particle preparation and structural characterization.** Hexagonal YPO$_4$:Eu$^{3+}$ single microcrystals were synthesized via a hydrothermal method. Briefly, 910 mg of yttrium nitrate hexahydrate (Y(NO$_3$)$_3$·6H$_2$O) and 56 mg of europium nitrate hexahydrate (Eu(NO$_3$)$_3$·6H$_2$O) were dissolved in 15 mL of deionized water and mixed thoroughly under magnetic stirring. Separately, 2385 mg of sodium carbonate (Na$_2$CO$_3$) and 288 mg of ammonium dihydrogen phosphate (NH$_4$H$_2$PO$_4$) were each dissolved in 15 mL of deionized water. The Na$_2$CO$_3$ solution was slowly added to the Y/Eu nitrate solution under continuous stirring, followed by an additional 30 minutes of mixing to ensure homogeneity. Subsequently, the NH$_4$H$_2$PO$_4$ solution was introduced dropwise into the mixture, and the resulting suspension was stirred for 1 hour. The final mixture was then transferred to a stainless-steel autoclave lined with polytetrafluoroethylene (PTFE) and heated at 180 °C for 9 hours. After natural cooling to room temperature, the resulting precipitate was collected by repeated centrifugation and washing with deionized water and absolute ethanol to remove residual ions. The purified product was dried at 80 °C for 12 hours to obtain the final microcrystalline powder. Structural analysis was performed by powder X-ray diffraction (XRD; Rigaku D/Max2550, 40 kV, 50 mA, Cu Kα radiation, $\lambda$ = 1.5406 Å) over a $2\theta$ range of 10°–60° with a step size of 8°/min (**Supplementary Fig. 1a**), enabling confirmation of the hexagonal crystal phase and compositional purity. Surface morphology was characterized using a field-emission scanning electron microscope (SEM; Hitachi S-4800, 3 kV), with samples sputter-coated with gold to minimize charging (**Supplementary Fig. 1b**). The spatial distribution of $Y^{3+}$ ions within the hexagonal lattice was visualized by generating the unit cell structure using VESTA software (**Fig. 2a**). For single-particle luminescence measurements, the microcrystalline powder was ultrasonically dispersed in ethanol at a concentration of 0.01 mg/mL to minimize aggregation. A droplet of this suspension was deposited onto a clean quartz substrate and dried under ambient conditions, resulting in a sparse, random in-plane distribution of isolated microcrystals (**Fig. 3a** and **Supplementary Fig. 11**) suitable for single-particle and ensemble optical characterization (**Fig. 4b**).

**Single-particle far-field polarized analysis.** The far-field polarization intensity for a single magnetic dipole transition was calculated using the analytical radiation formula for an oscillating magnetic dipole (**Eq. 6**)[63], implemented in Mathematica.

$$\vec{E}(\vec{r},t) = \frac{\mu_0 m_0 c k^2}{4\pi r} e^{j(\omega t - kr)}(\hat{r} \times \hat{m}) \tag{6}$$

In this equation, $\mu_0$ denotes the vacuum permeability, $m_0$ is the maximum magnetic dipole moment, $c$ is the speed of light, $k$ is the wave vector, $\omega$ is the angular frequency, and $r$ is the distance from the dipole to the observation point; $\hat{r}$ and $\hat{m}$ represent the unit vectors in the direction of the field point and the magnetic dipole, respectively. Given the sixfold rotational symmetry of the magnetic dipoles in a single microcrystal, the total far-field polarization intensity was determined as the incoherent sum of intensities from six equivalent dipoles. To resolve the 3D orientation of a single particle, a coordinate transformation was applied to the six magnetic dipoles using an appropriate rotation matrix ($\bar{\bar{R}}$). The transformed unit vectors ($\bar{\bar{R}}\hat{m}_{i=1,\cdots,6}$) were substituted into **Eq. 6** to analytically derive the polarization intensities for orthogonal linear polarizations, applicable to arbitrary 3D orientations of the particle (**Supplementary IV**).

**Single-particle polarized measurements.** To quantitatively investigate the polarized luminescence of single particles, a precision polarization-resolved microspectroscopy system was constructed (**Supplementary Fig. 7**). Excitation was achieved by generating a 395 nm linearly polarized laser beam via second-harmonic generation (SHG) of a titanium-sapphire femtosecond laser, which efficiently excites $Eu^{3+}$ ions to the $^5L_6$ state and induces fluorescence in $YPO_4:Eu^{3+}$ microcrystals. To eliminate any possible influence of excitation polarization on emission measurements, the linear polarization was converted to circular polarization using a half-wave plate (10RP02-48, Newport) and a quarter-wave plate (10RP04-48, Newport). In practice, owing to crystal phonon–mediated depolarization, the polarization state of non-resonant excitation does not affect the measured emission polarization of rare-earth dopants. The 395 nm beam was directed through a non-polarizing beam splitter, then focused onto individual microcrystals on a quartz substrate using a microscope objective (M Plan Apo 100x, NA = 0.55, Mitutoyo). The substrate, mounted on a three-axis nanopositioning stage, enabled precise spatial scanning. Emission and scattered excitation light from the sample were collected by the same objective and separated by a non-polarizing beam splitter, then passed through a 405 nm long-pass dichroic mirror (FL-007036, Semrock) to isolate the fluorescence signal. For polarization analysis, the fluorescence was transmitted through a rotatable half-wave plate (10RP52-1B, Newport) or quarter-wave plate (10RP54-1B, Newport) and a fixed linear polarizer (WP25M-VIS, Thorlabs), which was aligned parallel to the spectrometer slit (defined as 0°) to avoid polarization-dependent grating artifacts, before being delivered to a spectrometer (Shamrock SR-750-A, Andor) equipped with an EMCCD (DU970P-BVF, Andor). Spectral changes were recorded by rotating only the wave plate and analyzed by using the Poincaré sphere (PS) and polarization fitting (PF) methods (See **Supplementary Fig.7** for detail).

**Cell sample preparation and orthogonal polarization detection.** Mouse hepatocellular carcinoma cells (Hepa1-6) were cultured in Dulbecco's Modified Eagle Medium (DMEM) supplemented with 10% fetal bovine serum (FBS) at 37 °C in a 5% $CO_2$ atmosphere. For in vitro assays, approximately 50,000 cells were seeded onto pre-positioned glass coverslips in 12-well plates and allowed to adhere for 12 h. Subsequently, the culture medium was replaced with fresh medium containing microcrystals at a final concentration of 5 μg/mL, and incubation continued for an additional 4 h to enable cellular uptake. After incubation, the medium was aspirated, and cells were fixed with 4% paraformaldehyde (PFA) for 15 min at room temperature. Following fixation, samples were air-dried under sterile conditions and stored at –20 °C until further analysis. Bright-field image was performed at room temperature using a Leica DM2700 M microscope equipped with an MS60 digital camera (**Fig. 5g**). For single-view intracellular polarized detection, orthogonal polarization-resolved luminescence spectra were acquired by rotating the half-wave plate to 0° and 45°, corresponding to $I_{z'x'}$ and $I_{z'y'}$, respectively.

**Acknowledgements:**

We acknowledge the support from the National Key R&D Program of China (Grant Nos. 2023YFA1407100 and 2023YFB3508300), the National Natural Science Foundation of China (NSFC) (Grant Nos. 12474392, 12074303 and 52173290), and the Postdoctoral Fellowship Program of China Postdoctoral Science Foundation (Grant No. GZC20241349).

**Author contributions:**

P.L. and F.L. conceived the project and experiments. Y.G. and P.L. conducted the experiments. P.L., Y.Y., and F.L. performed the theoretical calculations and electromagnetic simulations with Mathematica®. P.L. and Y.G. analyzed the data. P.L. and F.L. supervised the project and wrote the manuscript with input from all the other authors.

**Competing interests:**

The authors declare no competing interests.

**Data and materials availability:**

All data needed to evaluate the conclusions in this manuscript are available in the main text or the supplementary materials.

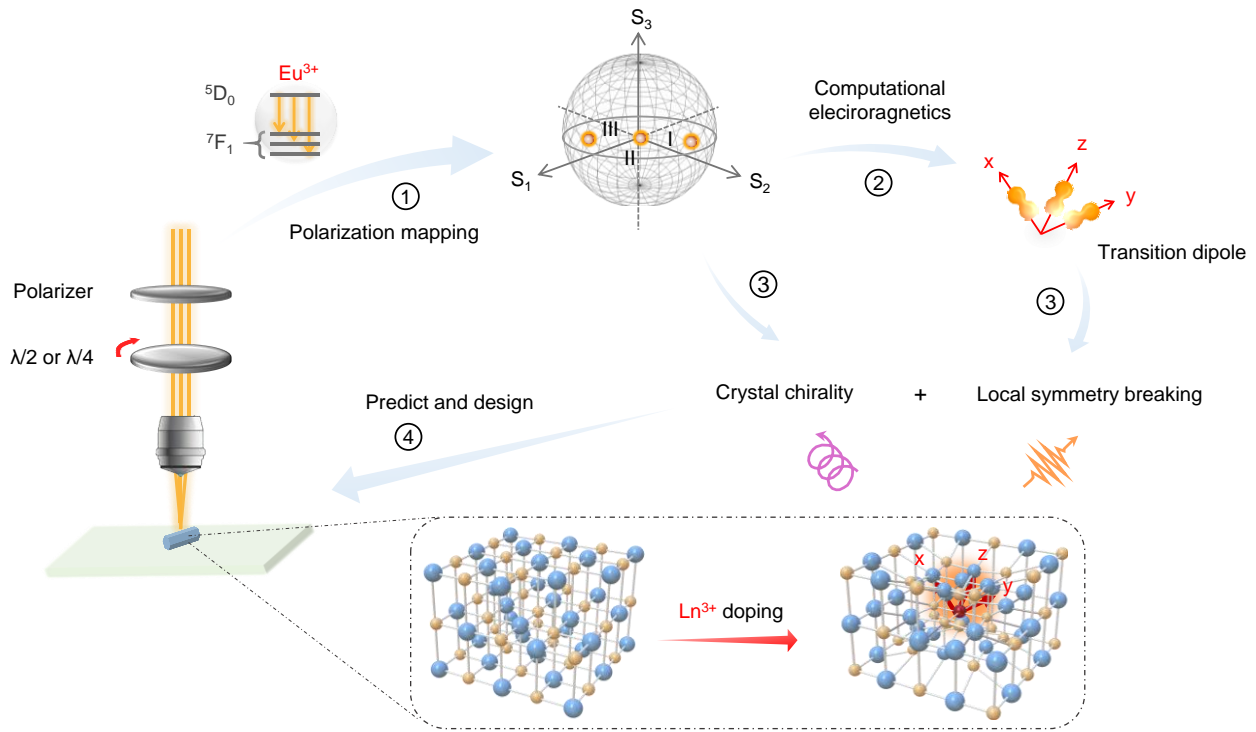

**Fig. 1 | Four-step model system for symmetry analysis of rare-earth doped single particle crystals: from polarization-resolved micro-spectroscopy to structure design.**

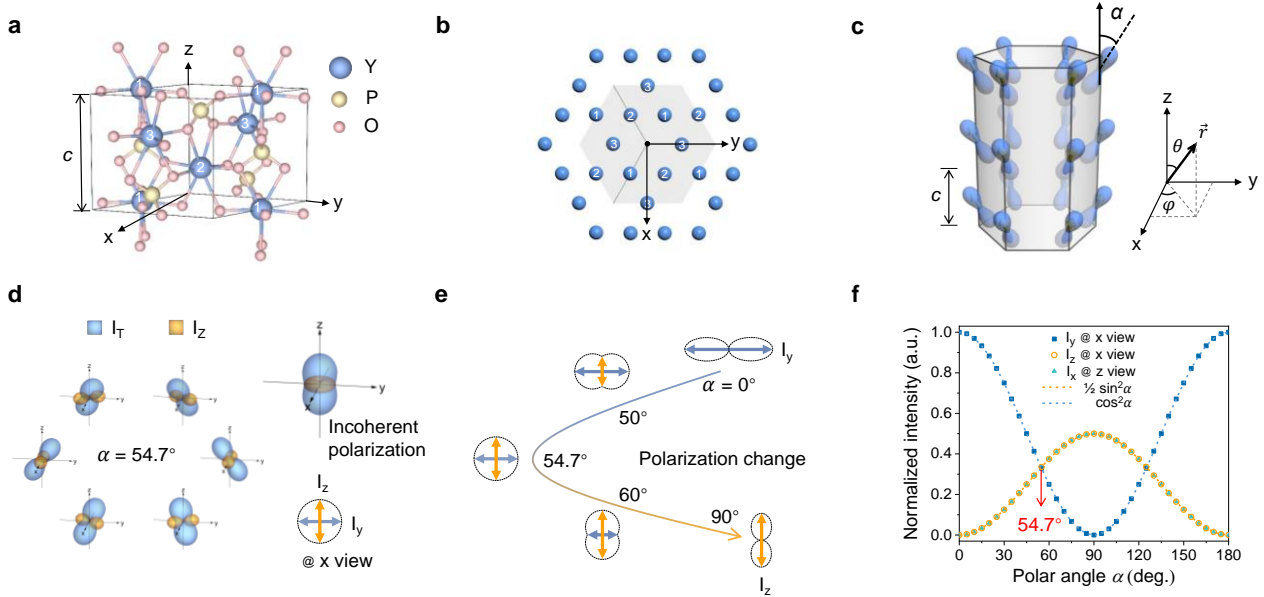

**Fig. 2 | Physical model of magnetic dipole optical transition and polar-angle-dependent far-field polarized intensities of $Eu^{3+}$ dopants in a single hexagonal $YPO_4$:$Eu^{3+}$ microcrystal. a** Unit cell of the hexagonal $YPO_4$ crystal, with the Cartesian x–y–z axes defined such that the z-axis is parallel to the crystallographic c-axis (the long axis of the crystal). **b** Projection of the $Y^{3+}$ (and $Eu^{3+}$ dopants) sublattice onto the x–y plane, revealing a Kagome lattice with sixfold rotational symmetry. **c** Model of the optical magnetic transition dipoles of $Eu^{3+}$ ions in the crystal. The polar angle $\alpha$ denotes the angle between the dipole and the c-axis. Inset: $\vec{r}$ indicates the position vector of the far-field observation point, specified by the spherical coordinates ($\theta$, $\varphi$). **d** Calculated far-field polarized intensity distributions at $\alpha = 54.7°$. The left subfigure shows the far-field radiation pattern of individual magnetic transition dipoles arranged with sixfold rotational symmetry; the top-right inset displays the total far-field intensity obtained by incoherently summing the emission from all six dipoles; the bottom-right inset illustrates the linearly polarized intensity components detected along the x-direction (the radial direction of the microcrystal). $I_z$ (yellow) and $I_T$ (blue, where $I_T = I_x + I_y$) represent the linearly polarized intensities with electric vectors parallel and perpendicular to the crystal c-axis, respectively. **e** The polarization variation detected along x-direction (the radial direction of the microcrystal) with varying $\alpha$. **f** Normalized far-field polarized intensity curves detected along different directions as a function of $\alpha$ for a single microcrystal, highlighting the angular dependence of the emission polarization.

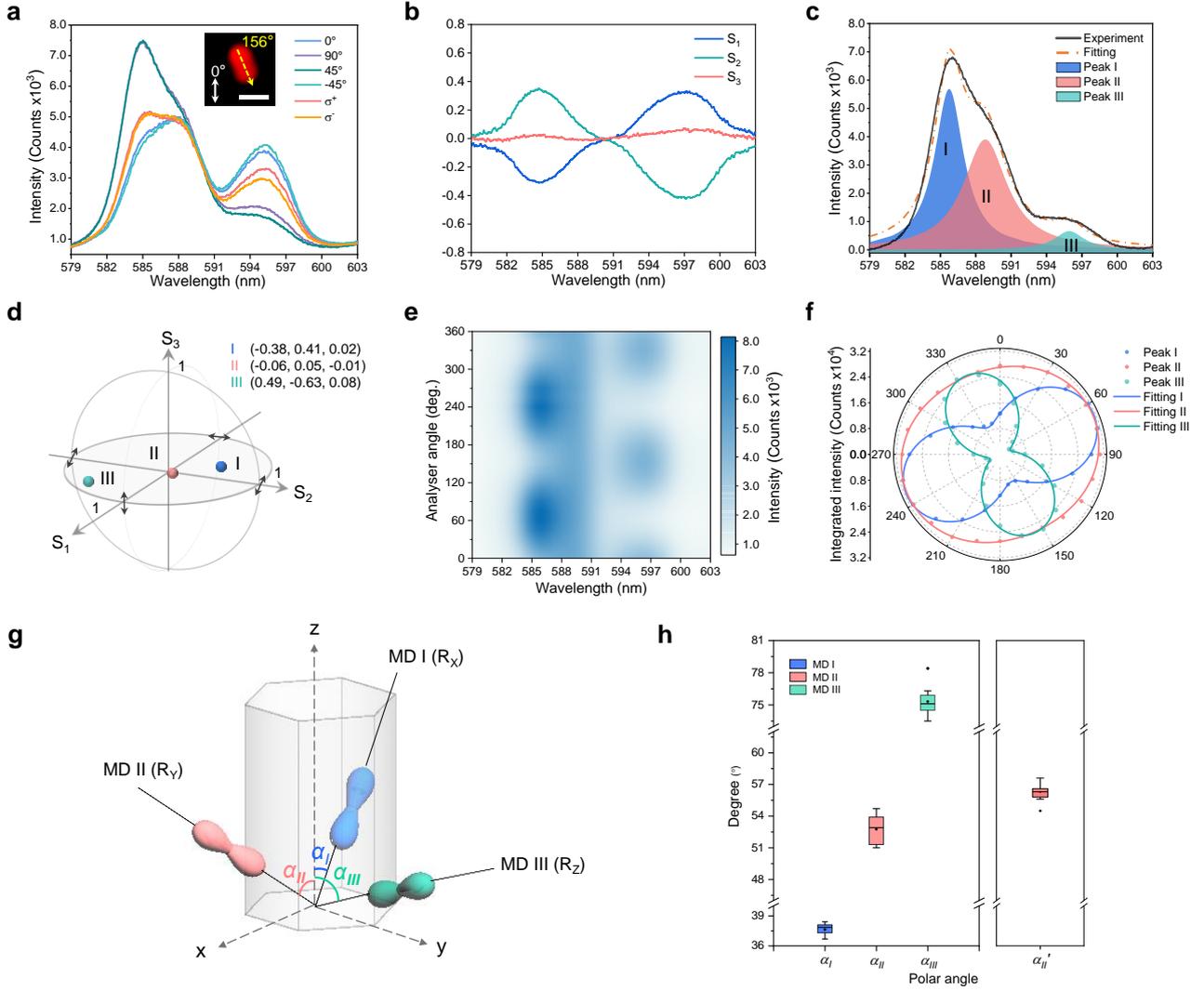

**Fig. 3 | Analysis of magnetic transition dipole orientation and point group symmetry of Eu$^{3+}$ dopants in single YPO$_4$ microcrystals. a** Polarization analysis using the Poincaré sphere (PS) method. Polarized photoluminescence (PL) spectra of the $^5D_0 \rightarrow {}^7F_1$ magnetic transition in a single microcrystal were recorded at the six Stokes bases. Inset: CCD image of the luminescent microcrystal lying on the substrate; its long axis (c-axis) is oriented at 156° relative to the spectrometer entrance slit (defined as 0°). Scale bar, 2 μm. **b** Stokes parameters $S_1$, $S_2$, and $S_3$ calculated from the polarized spectra. **c** Lorentzian fitting of the 45° polarized spectrum, revealing three emission peaks within the magnetic transition. **d** Coordinates ($S_1$, $S_2$, $S_3$) of the three emission peaks on the Poincaré sphere. **e** Intensity variation of the transition under continuous linear polarization detection. **f** Polarization fitting (PF) analysis of the three emission peaks. **g** Schematic representation of the three magnetic transition dipole orientations of Eu$^{3+}$ ions, each corresponding to one of the three emission peaks. The polar angles $\alpha_I$, $\alpha_{II}$, and $\alpha_{III}$ denote the angles between each dipole and the crystal c-axis (z-axis), respectively. **h**

Statistical analysis of $\alpha_I$, $\alpha_{II}$, and $\alpha_{III}$, as determined from linear polarization degrees (LDOPs) of the three peaks for ten single microcrystals using the polarization fitting (PF) method. Theoretical values $\alpha'_{II}$ (right) represent the expected dipole orientation for transition II, calculated from $\alpha_I$ and $\alpha_{III}$ in accordance with selection rules of the $D_2$ point group, and are shown for direct comparison with experimental $\alpha_{II}$ values.

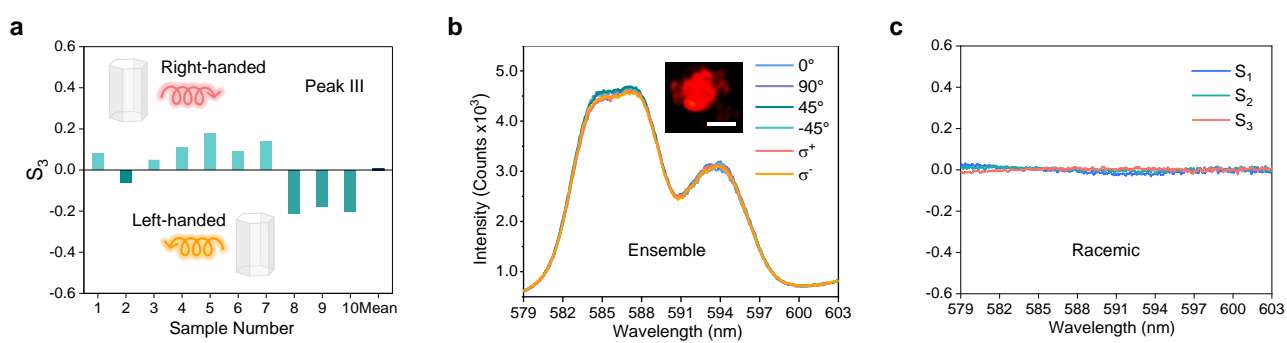

**Fig. 4 | Chiral analysis of single YPO$_4$:Eu$^{3+}$ microcrystal and its ensemble. a** Statistical distribution of S$_3$ values for peak III across ten randomly selected single microcrystals. **b** Polarization analysis of an ensemble of microcrystals using the PS method. Inset: CCD image of the luminescent ensemble. **c** Stokes parameters (S$_1$, S$_2$, S$_3$) for the ensemble. The results (S$_1$=S$_2$=S$_3$ =0) confirm the racemic nature and unpolarized emission of the ensemble. Scale bar, 4 μm.

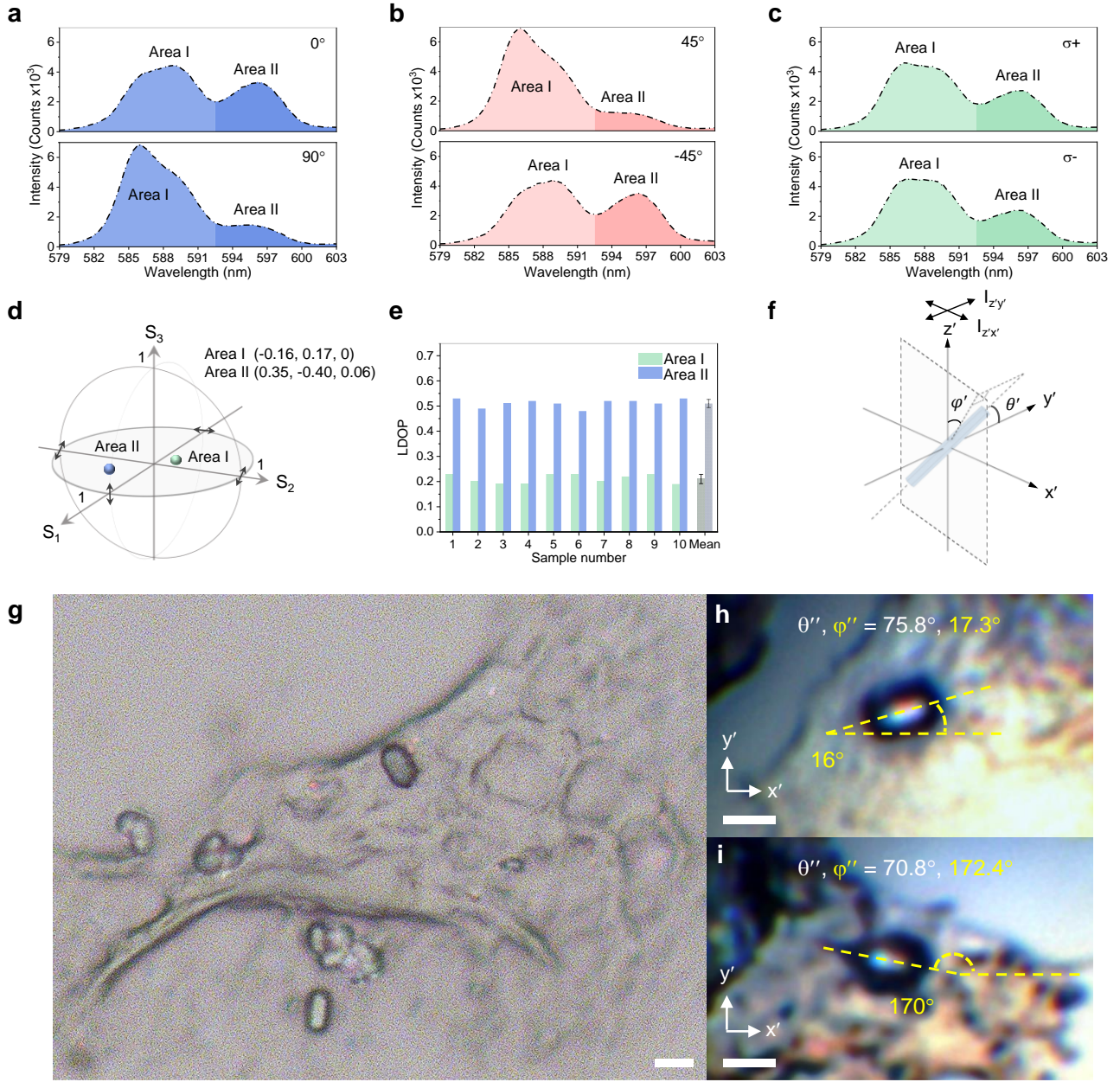

**Fig. 5 | Three-dimensional orientation determination of a single microcrystal based on polarized transition bands. a-c** Polarization analysis of the magnetic transition bands I (579–592.5 nm) and II (592.5–603 nm) in a single YPO$_4$;Eu$^{3+}$ microcrystal (sample no. 1) using the Poincaré sphere (PS) method. **d** Polarization coordinates of the two transition bands on the Poincaré sphere, indicating partial linear polarization near the equatorial plane. **e** Statistical analysis of the linear polarization degrees (LDOPs) for the two transition bands across ten single microcrystals; mean values and standard deviations are shown on the right, with the small deviations demonstrating the robustness of the LDOPs for these bands. **f** Schematic illustration of three-dimensional orientation determination for an individual micro- or nanocrystal. The polar angle $\theta'$ and azimuthal angle $\varphi'$ define the

orientation of the crystal c-axis within the Cartesian coordinate system. Orthogonally polarized intensities, $I_{z'x'}$ and $I_{z'y'}$, are measured along the $z'$-direction. **g** Bright-field image showing the spatial distribution of single microcrystals within and around a single cell. **h, i** Two examples of single-particle intracellular measurements. For these particles, the in-plane orientation angles are 16° and 170°, while the angles $\varphi''$ derived from orthogonal polarization spectra are 17.3° and 172.4°, and the corresponding in-plane tilt angles (90°-$\theta''$) are 14.2° and 19.2°, respectively. The orthogonal polarized spectra and the coordinate transformation relationships between ($\theta'$, $\varphi'$) and ($\theta''$, $\varphi''$) are detailed in **Supplementary Fig. 14**. Scale bars, 2 μm.